# Determination of the Spin Polarization of $RFe_2$ (R = Dy, Er, Y) by Point Contact Andreev Reflection

C. Morrison,[1,*] D. Wang[1,2], G. J. Bowden[1], R. C. C. Ward[3], and P. A. J. de Groot[1].

[1]*School of Physics and Astronomy, University of Southampton, Southampton SO17 1BJ, UK*

[2]*Department of Physics, National University of Defense Technology, Changsha 410073, Hunan, China*

[3]*Clarendon Laboratory, Oxford University, Oxford OX1 3PU, UK*

(Dated: 28$^{th}$ November, 2011)

Epitaxially grown intermetallic $RFe_2$ (R = Dy, Er, Y) thin films have been studied by point contact Andreev reflection. Spin polarization values were extracted by fitting normalized conductance curves for mechanical $Nb/RFe_2$ point contacts, using a modified Blonder-Tinkham-Klapwijk (BTK) model. Good agreement is found between this model and the experimentally obtained data. Extracted values of spin polarization, which are close to the spin polarization of Fe, reveal no variation with the rare earth component for the measured intermetallic compounds. This suggests that using this technique we probe the Fe sub-lattice, and that this lattice drives spintronic effects in these compounds.

---

[*] Current Address: School of Computer Science, University of Manchester, Manchester M13 9PL, UK

Exchange spring magnets have received much attention in recent years. Early work was related to potential applications in permanent magnets [1-3]. However, more recently, exchange spring media have been proposed for magnetic data storage [4-7]. Epitaxial rare earth – transition metal superlattices, comprising layers of $DyFe_2$, $ErFe_2$ and $YFe_2$, have been shown in previous work to be excellent model systems for the study of exchange springs [8,9]. They display a range of features, most notably exchange spring induced giant magnetoresistance (GMR) [10]. Rare earth intermetallics are known to demonstrate strong magneto-optic properties, which makes these materials relevant for novel, all-optical magnetic recording technologies [11].

Andreev reflection is observed at the interface between a normal metal and a superconductor [12]. For low bias voltages ($eV < \Delta$, where $\Delta$ is the superconducting band gap, $e$ is the electronic charge and $V$ is the voltage bias), electrons incident on the interface from the normal metal are retro-reflected as holes, while Cooper pairs propagate into the superconductor. A parallel hole conduction channel is formed, resulting in a doubling of the normal state conductance $dI/dV$ for zero bias voltage. If $eV > \Delta$ then formation of Cooper pairs at the superconductor side of the interface breaks down and normal conductance is observed.

Point contact Andreev reflection (PCAR) [13-15] is an excellent technique for determining the current spin polarization of a material, due to an imbalance of spin up and spin down conduction electron populations which suppresses the zero bias conductance for a superconductor/ferromagnet interface. Accurate knowledge of the spin polarization in magnetic materials is essential for applications in spin electronic or spintronic devices, given that the effectiveness of any device increases with spin polarization [16]. In this paper we report PCAR measurements using Nb point contacts to epitaxially grown $DyFe_2$, $ErFe_2$ and $YFe_2$ thin films.

The films were grown by molecular beam epitaxy (MBE) using the Balzers UMS 630 UHV facility at Oxford, following a procedure described by Bentall *et al.* [17]. A 100Å Nb buffer and a 20Å Fe seed layer were deposited onto an epi-prepared ($11\bar{2}0$) sapphire substrate. The $RFe_2$ material (cubic Laves structure) was grown in (110) orientation by co-deposition of elemental fluxes at a substrate temperature of 400ºC. The sample was then capped with a 100Å Y layer to prevent oxidization of the rare earth material.

A mechanical point contact [18] mounted on a differential screw system for micrometre control was used to make the measurements reported here. The differential conductance $G = dI/dV$ was measured as a function of applied voltage $V$ using standard AC lock-in techniques [19].

Figure 1 shows typical normalized differential conductance curves for point contacts of Nb to $DyFe_2$, $ErFe_2$ and $YFe_2$. The normalization is based on the relative value of $G(V)$ to the normal state conductance $G_n$, which can be approximated to $G(V)$ when $eV \gg \Delta$. In this regime the Nb tip can be viewed as a normal metal. Hence we plot $G(V)/G_n$ as a function of $V$, the applied voltage across the contact. The antisymmetric part of the conductance curves was removed [20], considering that Andreev reflection is even in *V*. Usual peculiar conductance features routinely observed in PCAR measurements were ruled out by careful selection of measured curves [21]. PCAR curves were obtained in the region of 5 to 50Ω: above 50Ω the contact becomes unstable and the data unreliable.

A simple model of superconducting to ferromagnetic metal point contacts assumes that the current flowing in the junction is split into two channels, polarized and unpolarized. Details of this model may be found in [15,22]. On using this relatively simple model, the spin polarization may be determined via $G(0)/G_n = 2(1-P_C)$. While this model may be used to provide an estimate of the contact spin polarization $P_C$, it does not include any consideration of temperature effects (*T*), and/or interfacial scattering effects (the parameter *Z*). As noted in the theoretical model of point contact Andreev reflection by Blonder, Tinkham and Klapwijk (BTK) [23], the *Z* parameter will result in an observed spin polarization that deviates from the intrinsic value for the material. Thus, any accurate measurement of the spin polarization $P_C$ must take into account interfacial scattering. In addition, a proper description of real point contacts should consider also the effect of the spreading resistance ($R_s$) [24, 25].

For a more thorough analysis of the data a modified model which includes both *Z* and temperature *T* must be used. In this paper we follow the methodology of Mazin *et al.* [26], assuming purely ballistic contacts, although our contacts are in the diffusive regime, which is evident from the measured second derivative (not shown here), $d^2I/dV^2$, curves. Indeed, the ballistic formulation can provide a good fit for any data, ballistic or diffusive, as long as the precise value of *Z* is not required, as concluded by Woods *et al.* [24]. Following a similar argument to the simple model of Eq. (1), the current is split into two channels, polarized and unpolarized, thus $I = (1 - P_C)I_N + I_P$. The expressions for the Andreev and normal reflection coefficients $A(E)$ and $B(E)$ in the ballistic nonmagnetic and half-metallic regimes, taken from reference [26], are then used to obtain an expression for the total current across the junction,

$$I = 2e\alpha N v_F \int_{-\infty}^{\infty} [f(E-eV,T) - f(E,T)] \times [1 + A(E) - B(E)]dE \qquad (1)$$

Here $\alpha$ is the effective cross-sectional area of the contact, $f$ is the Fermi-Dirac distribution function, $N$ is the spin dependent density of states at the Fermi energy and $v_F$ is the Fermi velocity.

Numerical fitting of experimental data to equation 1, including the effect of $R_s$ [25], can be seen in Fig. 1 with the extracted parameters ($T$, $R_s$, $\Delta$, $Z$ and $P_C$). The quality of each individual least-squares fitting procedure was evaluated using a chi-squared analysis, which involved evaluating the value of chi-squared as a trial value of $P_C$ was varied as a fixed parameter in the fit [27]. A clear, sharp minimum in the chi-squared curves as a function of $P_{trial}$ indicates confidence in the value of $P_C$ obtained from the fit. On averaging conductance curves measured at different contact resistances, the spin polarization is determined to be $P_C = (40\pm1)\%$ for DyFe$_2$, $P_C = (40\pm3)\%$ for YFe$_2$ and $P_C = (39\pm3)\%$ for ErFe$_2$. The error in these values was determined by averaging over multiple analyses of separate sets of raw data for each material. All measurements with large $Z$ are not used to get the spin polarization $P_C$, so effectively we have used only 4 free parameters for the fitting. What should be noted here is that the fitting temperature $T$ is only a measure of the broadening of the PCAR spectra [24, 27]. Due to various thermal and non-thermal contributions to spectrum broadening, such as local heating and pair breaking effects [24, 28], the fitted $T$ could be higher than the actual temperature of the contact.

The value of spin polarization obtained for all three compounds is very similar to the value for Fe, obtained via PCAR by previous authors [15, 29]. This suggests that the rare earth element has very little effect on the spin polarization of the compound. At first sight this is surprising given the different structures of the cubic Laves $RFe_2$ and bcc elemental Fe. However in $YFe_2$ the 4d moment on the Y site is small ($0.44\mu_B$) [30], while the moment on the Fe site has been obtained from neutron scattering as $2.3\pm0.3\mu_B$ [31], close to that of elemental Fe ($2.2\mu_B$). Thus conduction band occupation at the Fermi surface, which is important for the determination of spin polarization $P_C$, is likely to be very similar for both bcc-Fe and $YFe_2$. In $DyFe_2$, magnetism from the Dy sites is due to the 5d moments, driven primarily by the Fe sublattice via 3d-5d hybridization [32]. The calculated 5d moment in $DyFe_2$ is $0.53\mu_B$, antiparallel to the moment on the Fe site. This is relatively small compared to the total calculated conduction electron magnetic moment of $DyFe_2$ ($3.08\mu_B$) from Brooks *et al.*, suggesting that the rare earth site plays a relatively small role in magnetic behaviour near the Fermi level. The same conclusion holds for $ErFe_2$ [32]. Experimentally, anomalous Hall effect measurements performed on $ErFe_2/YFe_2$ multilayers suggest that the Fe moments are dominant in determining electron transport [33], further supporting the conclusion that in these PCAR measurements, which essentially measure the current spin polarization, we are primarily probing the Fe sublattice.

Domain wall magnetoresistance at low temperature has been observed in DyFe$_2$/YFe$_2$ exchange-spring superlattices [10]. It was found that a giant magnetoresistance ratio of 32% can be achieved for narrow domain wall width, $\delta_w \sim 20$Å. The Levy-Zhang formula [34] was used by Gordeev *et al.* to account for additional resistance derived from domain wall scattering,

$$\frac{\Delta\rho_w}{\rho_0} = \left(\frac{\pi\hbar^2 k_F}{4mJ\delta_w}\right)^2 \frac{(1-\alpha^2)}{5\alpha} \qquad (2)$$

where $J$ is the magnetic exchange constant, $k_F$ is the Fermi wave vector, $m$ is the effective mass of the electron and $\alpha = \rho_0^{\uparrow}/\rho_0^{\downarrow}$ is the spin asymmetry parameter between the spin-down and spin-up resistivities. If typical values for RFe$_2$ intermetallic compounds are adopted, $k_F = 2.2$Å$^{-1}$, $J = 0.4$eV, and assuming the free electron mass, $\Delta\rho_w/\rho_0 = 32\%$ for a $\delta_w = 20$Å domain wall corresponds to a spin polarization $P_C = |\alpha-1|/(\alpha+1) = 37\%$, which is very close to the values of $P_C$ determined from our PCAR measurements. This consistency is really striking, considering the approximations used in the calculation.

In conclusion, point contact Andreev reflection measurements have been performed on three epitaxially grown RFe$_2$ thin films. Measurements were repeated for a number of contact resistances, corresponding to differing contact radii. Fitting to a modified Blonder-Tinkham-Klapwijk model using 5 free parameters, including the spreading resistance contribution, allowed the intrinsic value of the spin polarization $P_C$ to be extracted for all three compounds. The extracted $P_C$ is close to the spin polarization of Fe, suggesting that the conduction of RFe$_2$ compounds is mainly mediated by 3d electrons.

Financial support from EPSRC UK, under Grant No. GR/S95824, is gratefully acknowledged. D.W. acknowledges financial support from the Chinese and the UK Governments under the Scholarships for Excellence.

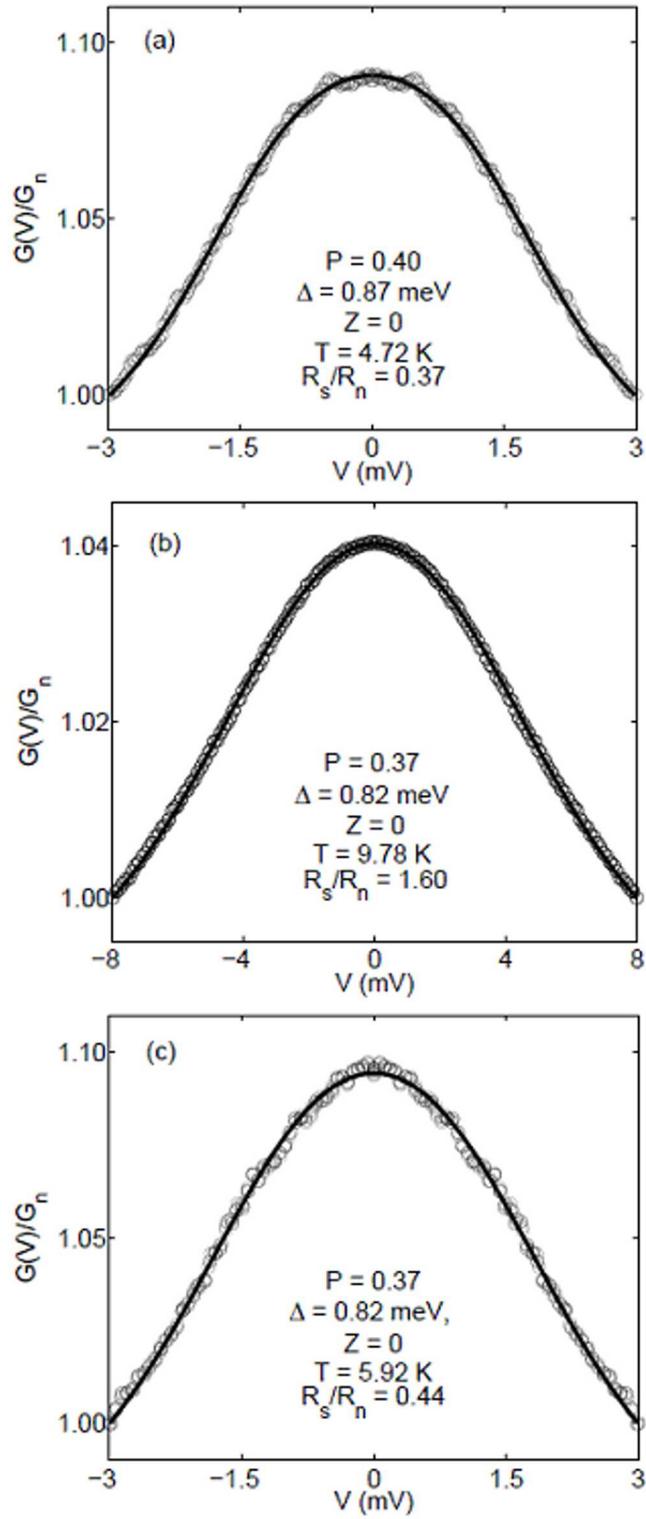

Figure 1: Normalized conductance and best fit for (a) a 10Ω $DyFe_2$ contact, (b) a 33Ω $ErFe_2$ contact and (c) a 30Ω $YFe_2$ contact.